\documentclass[twocolumn,prb,aps,floatfix,showpacs]{revtex4}
\usepackage{amssymb}
\usepackage{epsfig}

\usepackage{amsmath}
\usepackage{graphicx}
\usepackage{dcolumn}
\usepackage{bm}

\begin{document}
\title{Valley Dependent Optoelectronics from Inversion Symmetry Breaking}

\author{Wang Yao}
\author{Di Xiao}
\author{Qian Niu}
\affiliation{Department of Physics, The University of Texas, Austin,
  TX 78712-0264}

\begin{abstract}
Inversion symmetry breaking allows contrasted circular dichroism in
different $k$-space regions, which takes the extreme form of optical
selection rules for interband transitions at high symmetry points.
In materials where band-edges occur at noncentral valleys, this
enables valley dependent interplay of electrons with light of
different circular polarizations, in analogy to spin dependent
optical activities in semiconductors. This discovery is in perfect
harmony with the previous finding of valley contrasted Bloch band
features of orbital magnetic moment and Berry curvatures from
inversion symmetry breaking [Phys. Rev. Lett. {\bf 99}, 236809
(2007)]. A universal connection is revealed between the $k$-resolved
optical oscillator strength of interband transitions, the orbital
magnetic moment and the Berry curvatures, which also provides a
principle for optical measurement of orbital magnetization and
intrinsic anomalous Hall conductivity in ferromagnetic systems. The
general physics is demonstrated in graphene where inversion symmetry
breaking leads to valley contrasted optical selection rule for
interband transitions. We discuss graphene based valley
optoelectronics applications where light polarization information
can be interconverted with electronic information.
\end{abstract}

\pacs{78.67.-n, 81.05.Uw, 78.20.Ls, 85.60.-q} \maketitle


\section{Introduction}

In atoms, optical transition selection rules are determined by
orbital magnetic moments of the atomic levels. Bloch bands in solids
can inherit these rules from their parent atomic orbits. A well
known example is the optical transition between $s$-like conduction
and $p$-like valance bands in semiconductors. Strong spin-orbit
coupling in the $p$ bands mixes orbital moment with spin moment, and
as a result, spin degree of freedom distinguishes two groups of
electrons in their response to light with opposite circular
polarizations. This forms the basis for spin optoelectronics
applications in semiconductors.~\cite{Ivchenko1,Ivchenko2,wolf2001}

In this paper, we explore an entirely different origin of selection
rules and circular dichroism for optical interband transitions in
solids. In addition to the intracellular current circulation of the
parent atomic orbits, orbital magnetic moment of Bloch electrons has
a contribution from intercellular current circulation governed by
bulk symmetry properties. We show that this contribution is also
tied to optical circular dichroism. When inversion symmetry is
broken, contrasted circular dichroism is allowed in different
regions of the Brillouin zone, which takes the extreme form of
optical transition selection rules at high symmetry points.

We illustrate this general physics in graphene, the monolayer carbon
honeycomb lattice recently realized in free-standing
forms.~\cite{novoselov2004,chiral_QHE1,chiral_QHE2,geim2007} The
isolated graphene crystallite is a zero-gap semiconductor: the
conduction and valence bands conically touch each other, forming two
inequivalent valleys at the corners of the first Brillouin zone.
Inversion symmetry breaking is being exploited as a powerful
approach towards bandgap
engineering,~\cite{ohta2006,bilayer_theory1,bilayer_theory2,bilayer_theory3,grapheneGap_Lanzara,Giovannetti2007}
motivated by the need of semiconductor energy gap for graphene based
logic devices.~\cite{geim2007} In graphene single layer, inversion
symmetry is broken when the two sublattices become inequivalent.
This effect is generally expected in epitaxially grown
graphene,~\cite{geim2007} where staggered sublattice potential can
arise either directly from the substrate such as
BN,~\cite{Giovannetti2007} or from a carbon buffer monolayer
covalently bonded to SiC
substrate.~\cite{grapheneGap_Lanzara,Mattausch2007} Recent angular
resolved photo emission spectroscopy has identified mid-infrared
bandgap in graphene epitaxially grown on
SiC,~\cite{grapheneGap_Lanzara} attributed to this mechanism. In
graphene bilayer, experiments and theories have revealed an energy
gap continuously tunable from zero to mid-infrared by an interlayer
gate bias that breaks the inversion
symmetry.~\cite{ohta2006,bilayer_theory1,bilayer_theory2,bilayer_theory3}
We find that at the center of the two inequivalent valleys,
interband transitions couple exclusively to optical field of
opposite circular polarizations.

Besides graphene, many conventional semiconductor materials have
noncentral $k$-space valleys, e.g. Si and AlAs. Valley-based
electronics applications have recently attracted great interests
where this extra degree of freedom is suggested as an information
carrier.~\cite{Shayegan_valley2,Beenakker_valleytronics1,Beenakker_valleytronics2}
In conjugation with the progressively achieved advances in
piezoelectric control and magnetic control of valley degree of
freedom,~\cite{Shayegan_valley3,Shayegan_valley4,Shayegan_valley5}
valley contrasted optical circular dichroism from inversion symmetry
breaking creates an entirely new possibility to optically address
the valley physics. This may form the basis of valley based
optoelectronics applications in graphene and conventional
semiconductors, in direct analogy to the well developed spin based
optoelectronics.~\cite{Ivchenko1,Ivchenko2,wolf2001}

This paper is organized as follows. In section \ref{sec_general}, a
universal connection is revealed between the $k$-resolved optical
oscillator strength of interband transitions, and the band
properties of orbital magnetic moment and Berry curvatures. We show
how inversion symmetry breaking allows contrasted circular dichroism
for interband transitions in different regions of the Brillouin
zone, which becomes a rigorous optical selection rule at high
symmetry points. In section \ref{section_rule}, we demonstrate the
valley contrasted optical selection rule in two qualitatively
different graphene systems with broken inversion symmetry, i.e.
single layer with staggered sublattice potential and bilayer with an
interlayer gate bias. In section \ref{section_application}, we
discuss graphene based valley optoelectronics applications where
light polarization information can be interconverted with electronic
information. In appendix \ref{appen_sum}, we give two circular
dichroic sum rules for optical measurement of orbital magnetization
and intrinsic anomalous Hall conductivity in ferromagnetic systems,
which are direct consequence of the connection between interband
optical transitions and the band topological properties discussed in
section \ref{sec_general}.

\section{General theory} \label{sec_general}

We consider a non-degenerate band $c$ for which the intracellular
circulation currents from the parent atomic orbit is assumed
absent.~\cite{nondegenerate} Orbital magnetic moment contributed by
the intercellular current circulation can be given from the $\bm{k
\cdot p}$ analysis~\cite{Yafet_gfactor}
\begin{equation}
\bm{m}(\bm k) \equiv -i \frac{e \hbar }{2 m_e^2} \sum_{i \neq c}
\frac{\bm{\mathcal{P}}^{ci}(\bm{k}) \times
\bm{\mathcal{P}}^{ic}(\bm{k})}{ \varepsilon_i(\bm{k}) -
\varepsilon_c(\bm{k})}~. \label{obitalmoment}
\end{equation}
In the above, $ \mathcal{P}^{ci}_{\alpha}(\bm{k}) \equiv \langle
u_{c,\bm{k}} | \hat{p}_{\alpha} | u_{i,\bm{k}} \rangle$ is the
interband matrix element of the canonical momentum operator,
$\varepsilon(\bm{k})$ is the band dispersion, and $|u_{\bm
k}\rangle$ the periodical part of the Bloch
function.~\cite{Yafet_gfactor,chang1996} It is worth noting that
this intercellular current circulation is responsible for the
anomalous $g$-factor of electron in semiconductors: the two Bloch
states with opposite spin at $\Gamma$ point form a Kramer's pair
with opposite $\bm{m}$ which renormalize the spin Zeeman
energy.~\cite{Yafet_gfactor}

Now we consider the simplest situation of a two band model, where
$\bm{m}(\bm{k})$ will have identical value in the upper and lower
bands as evident from Eq.~(\ref{obitalmoment}). The projection of
$\bm{m}(\bm{k})$ along the light propagation direction ($\hat{\bm
z}$) can be expressed as
\begin{equation}
-2 \frac{ \bm{m} (\bm k) \bm{\cdot} \hat{\bm z} }{\mu_B} =
\frac{|\mathcal{P}_+(\bm k)|^2 - |\mathcal{P}_-(\bm k)|^2}{m_e
\left(\varepsilon_c(\bm{k}) - \varepsilon_v(\bm{k}) \right) }
\label{difference},
\end{equation}
where the right hand side is the difference in $k$-resolved
oscillator strength of $\sigma+$ and $\sigma-$ circular
polarizations.~\cite{Yu_Cardona} $ \mathcal{P}_{\pm} \equiv
\mathcal{P}^{cv}_x \pm i \mathcal{P}^{cv}_y $, and $\mu_B \equiv
\frac{e \hbar}{2 m_e}$ is the Bohr magneton of free electron.
Further, with the equality for the polarization averaged oscillator
strength
\begin{equation}
\frac{|\mathcal{P}_+(\bm k)|^2 + |\mathcal{P}_-(\bm k)|^2 }{2 m_e
\left( \varepsilon _{c}\left( \bm{k} \right) -\varepsilon _{v}\left(
\bm{k}\right) \right) } = m_e {\rm Tr} \left [ \frac{1}{2}
\frac{\partial ^{2}\varepsilon _{c}\left( \bm{k}\right) }{\hbar^2
\partial k_{\alpha}\partial k_{\beta}} \right ] \label{sum},
\end{equation}
we find, for the interband transition at a $k$-space point, the
degree of circular polarization is given by
\begin{equation}
\eta(\bm k) \equiv \frac{|\mathcal{P}_+(\bm k)|^2 -
|\mathcal{P}_-(\bm k)|^2}{|\mathcal{P}_+(\bm k)|^2 +
|\mathcal{P}_-(\bm k)|^2} = - \frac{\bm{m}(\bm k) \bm{\cdot}
 \hat{\bm z} }{\mu_B^{\ast}(\bm k)}, \label{ratio}
\end{equation}
where $\mu_B^{\ast}(\bm k)$ is the {\it effective} Bohr magneton
with the bare electron mass replaced by the isotropic part of the
{\it effective mass}. Generalization of these relations to
many-bands is straightforward where the contribution from each pair
of bands to the $k$-resolved oscillator strength and orbital
magnetic moment assumes a similar relation.

Such connections open up the possibility to engineer optical
circular dichroism in given bands through intercellular circulation
currents determined by bulk symmetry properties. In the presence of
time reversal symmetry, two Bloch states with opposite crystal
momentum form a Kramer's pair with opposite orbital moment, and
hence the overall circular dichroism vanishes. On the other hand,
inversion symmetry dictates that a Bloch state at $\bm{k}$ also has
a counterpart at $-\bm{k}$ with identical orbital moment. Thus,
inversion symmetry breaking is a necessary condition for contrasted
circular dichroism in different regions of the Brillouin zone. At
high symmetry points where the Bloch states are invariant under a
$q$-fold discrete rotation about the light propagation direction:
$\mathcal {R}(\frac{2\pi}{q},\hat{\bm z}) |\psi_{c (v),\bm k}\rangle
= e^{-i \frac{2\pi}{q} l_{c (v)}} |\psi_{c (v),\bm k}\rangle$, an
azimuthal selection rule $l_v\pm1=l_c+qN$ is expected for interband
transitions by light of $\sigma\pm$ circular polarization.

Berry curvature is another property that reflects handedness of
Bloch electrons and it always accompanies the intercellular current
circulations in the Bloch band~\cite{chang1996}
\begin{equation}
\bm{\Omega}(\bm k) \equiv i \frac{\hbar^2}{m_e^2} \sum_{i \neq c}
\frac{\bm{\mathcal{P}}^{ci}(\bm{k}) \times
\bm{\mathcal{P}}^{ic}(\bm{k})}{ \left(\varepsilon_c(\bm{k}) -
\varepsilon_i(\bm{k}) \right)^2}~. \label{curvature}
\end{equation}
While moving in an in-plane electric field, the carriers acquire an
anomalous velocity in the transverse direction proportional to the
Berry curvature, the charge and the field, giving rise to the Hall
effect.~\cite{chang1996,QHE_TKNN} For the two band model discussed
above, we simply find
\begin{equation}
\eta(\bm k) = - \frac{\bm{m}(\bm k) \bm{\cdot}
 \hat{\bm z} }{\mu_B^{\ast}(\bm k)} = - \frac{\bm{\Omega}(\bm k) \bm{\cdot}
 \hat{\bm z} }{\mu_B^{\ast}(\bm k)} (\varepsilon_c(\bm{k}) -
\varepsilon_i(\bm{k}) ) \frac{e}{2 \hbar} , \label{curvature2}
\end{equation}
where $\bm{m}(\bm{k})$ and $\bm{\Omega}(\bm{k})$ both stands for
their value in the upper band (we note that $\bm{\Omega}(\bm{k})$
will have opposite value in the upper and lower bands as evident
from Eq.~(\ref{curvature})). Hence, valley contrasted optical
circular dichroism is also generally accompanied by a valley
contrasted contribution to the Hall conductivity. This makes
possible optoelectronic schemes implementing this topological
transport phenomena, as will be discussed in
section~\ref{section_application}.

In ferromagnetic systems, the universal connection between $\eta(\bm
k)$, $\bm{m} (\bm k)$ and $\bm{\Omega}(\bm k)$ directly leads to the
dichroic sum rules for optical measurement of orbital magnetization
and intrinsic anomalous Hall conductivity, as given in
appendix~\ref{appen_sum}.

\section{Valley contrasted optical selection rule in graphene}
\label{section_rule}

It has been shown by previous studies that a finite bandgap opens in
graphene as a generic consequence of inversion symmetry
breaking.~\cite{grapheneGap_Lanzara,Giovannetti2007,ohta2006,bilayer_theory1,bilayer_theory2,bilayer_theory3}
Optical interband transitions can then be addressed in a similar way
as that in conventional direct bandgap semiconductors. The bandedges
of both the conduction and valance bands occur at the Dirac points
which have the three-fold discrete rotational symmetry. We first
demonstrate the valley contrasting selection rules in single layer
graphene with staggered sublattice potential. More complex optical
activities in biased graphene bilayers which enables additional
optical control possibilities is presented next.

\begin{figure}
\includegraphics[width=7cm, bb=100 408 380 679]{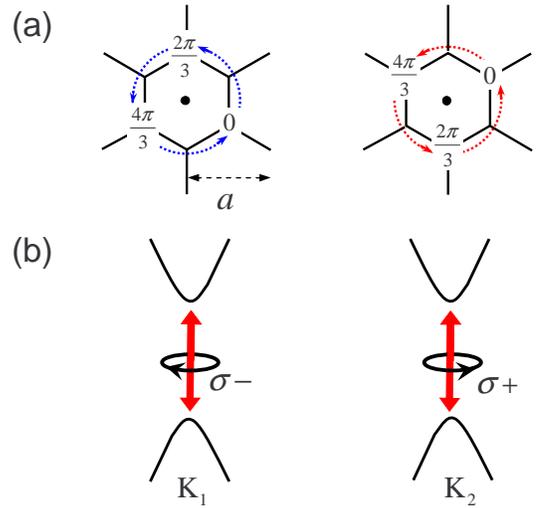}
\caption{(a) Left (right): phase winding of the conduction (valance)
band Bloch function at $\bm{K}_1=-\frac{4\pi}{3a} \hat{\bm x}$,
showing the intercellular current circulations. (b) Valley optical
selection rules: $\sigma+$ ($\sigma-$) circularly polarized light
couples only to bandedge transitions in valley $\rm{K}_2$
($\rm{K}_1$)} \label{phase}
\end{figure}

The active $\pi$ bands in graphene originate from $p_z$ atomic orbit
with zero moment along the normal direction of the plane. The
extremely weak spin-orbit coupling can be
neglected.~\cite{SO_graphene1,SO_graphene2} In the tight binding
approximation, graphene single layer with staggered sublattice
potential can be modeled with nearest neighbor hopping energy $t$
and a site energy difference between sublattices
$\Delta$~\cite{grapheneGap_Lanzara,Giovannetti2007}
\begin{equation}
\hat{H}\left( \bm{k}\right) =\left[
\begin{array}{cc}
\Delta/2 & V\left( \bm{k}\right) \\
V^{\ast }\left( \bm{k}\right) & -\Delta/2
\end{array}
\right] \label{singlelayer}.
\end{equation}
$V (\bm{k} ) = -t\left( e^{i\bm{k} \bm{\cdot} \bm{d}_{1}}+e^{i \bm{
k} \bm{\cdot} \bm{d}_{2}}+e^{i\bm{k} \bm{\cdot} \bm{d}_{3}}\right)$
where $\bm{d}_{1,2}=\frac{a}{2\sqrt{3}} \hat{\bm x} \pm
\frac{a}{2}\hat{\bm y}, \bm{d}_3= -\frac{a}{\sqrt{3}}\hat{\bm x} $
with $a$ being the lattice constant. The two component wavefunction
represents the amplitude on sublattice A and B respectively. Without
losing generality, we assume $\Delta>0$, i.e. sublattice A has a
larger on-site energy. Eq.~(\ref{singlelayer}) has the solutions of
a positive energy band (conduction) with dispersion
$\varepsilon_c(\bm k)$ and a negative energy band (valance) with
dispersion $\varepsilon_v(\bm k)=-\varepsilon_c(\bm k)$, separated
by an energy gap of $\Delta$. $\varepsilon_c(\bm k)$ has two valleys
centered at the Dirac points $\bm{K}_{1,2} = \mp \frac{4\pi}{3a}
\hat{\bm x}$ for which we introduce the valley index $\tau_z=\pm$.
Near the Dirac points
\begin{equation}
|\mathcal{P}_{\pm}(\bm k)|^2  = m_e^2  v_0^2 (1 \mp \tau_z \cos
\theta)^2 \label{probability},
\end{equation}
where $v_0=\frac{\sqrt{3} a t}{2 \hbar}$ is the Fermi-Dirac velocity
in graphene and $\cos \theta = \Delta/(\varepsilon_c(\bm
k)-\varepsilon_v(\bm k))$. At the bottom of valleys where
$\varepsilon_c(\bm k)-\varepsilon_v(\bm k)\simeq \Delta$, optical
transition is strongest: $|\mathcal{P}|^2/m_e \sim 20$~eV,
comparable to that for the transition between $\Gamma_6$ conduction
and $\Gamma_8$ valance bands in GaAs. Most significantly, there is
nearly perfect optical selection rule: $\sigma+$ ($\sigma-$)
circularly polarized light couples only to bandedge transitions in
valley $\rm{K}_2$ ($\rm{K}_1$) [Fig.\ref{phase}]. The rule is exact
at the Dirac points where the conduction (valance) band state is
constructed entirely from the orbits on sublattice A (B) and we have
$l_c=\tau_z$ ($l_v=-\tau_z$) under the 3-fold discrete rotation (see
Fig.\ref{phase}).~\cite{moment} Far away from the Dirac points
$\varepsilon_c(\bm k) -\varepsilon_v(\bm k) \gg \Delta$, circular
dichroism disappears as in the isolated graphene sheet and we
reproduce the constant high frequency optical conductivity found in
Ref.~\onlinecite{GrapheneConductivity1,GrapheneConductivity2}.

\begin{figure}
\includegraphics[width=7cm]{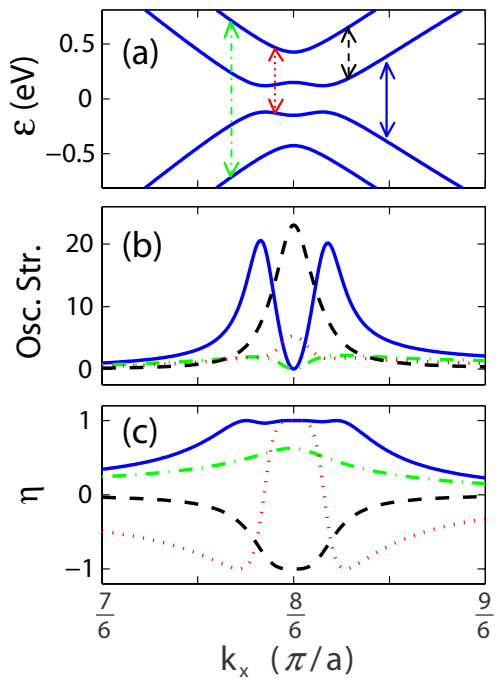}
\caption{(Color online) Optical properties of interband transitions
in biased graphene bilayer. Energy dispersions are given in (a).
$k$-resolved oscillator strength averaged over polarization (b), and
degrees of circular polarization (c) are shown for valley
$\rm{K}_2$. The values in valley $\rm{K}_1$ can be obtained by
noting that the oscillator strength is even while the degrees of
circular polarization is odd function of $\bm{k}$. Different line
style and color are used for transitions between different pairs of
bands as indicated by arrowed lines in (a). The parameters used are
$t = 2.82$ eV, $\Delta = 0.3$ eV, and $t_\perp = 0.4$ eV.}
\label{optical}
\end{figure}

In graphene bilayer with Bernal stacking, the A sublattice of the
upper layer sits on top of the B sublattice of the lower layer. The
band properties are well described by the tight-binding
approximation with an intra-layer nearest neighbor hopping $t$, an
interlayer nearest neighbor hopping $t_{\perp}$, and an interlayer
bias
$\Delta$,~\cite{bilayer_theory1,bilayer_theory2,bilayer_theory3}
\begin{equation}
H\left( \mathbf{k}\right) =\left[
\begin{array}{cccc}
\frac{\Delta }{2} & V\left( \mathbf{k}\right) & 0 & 0 \\
V^{\ast }\left( \mathbf{k}\right) & \frac{\Delta }{2} & t_{\perp } & 0 \\
0 & t_{\perp } & -\frac{\Delta }{2} & V\left( \mathbf{k}\right) \\
0 & 0 & V^{\ast }\left( \mathbf{k}\right) & -\frac{\Delta }{2}
\end{array}
\right]~.
\end{equation}
The obtained band structures (see Fig.~\ref{optical}(a)) agree well
with the measurement using angular resolved photo emission
spectroscopy.~\cite{ohta2006} The bilayer graphene has two positive
energy bands (conduction) and two negative energy bands (valance).
$k$-resolved oscillator strength and the degree of circular
polarization are shown for interband transitions between the two
conduction and the two valance bands. For the transitions between
the lower conduction band and the higher valance band, a nearly
perfect selection rule is obtained near the Dirac points where
valley $\rm{K}_1$ ($\rm{K}_2$) favors $\sigma-$ ($\sigma+$)
polarized light, similar to that in the graphene single layer with
staggered sublattice potential. A distinct feature for this
interband transition is the distribution of oscillator strength
sharply concentrated at the band-edge of the Mexican hat like energy
dispersion, in concert with the distribution of orbital magnetic
moment and Berry curvatures found in Ref.~\onlinecite{XYN2007}.
Interestingly, the transition between the two conduction bands also
has a perfect selection rule but with opposite chirality in the
vicinity of the Dirac points. The richer band structures and optical
activities allow more optical control possibilities in bilayer
graphene as discussed in section \ref{section_application}.

In the above, we have assumed the value of $\Delta>0$. In the
opposite case where $\Delta<0$, the results remains the same except
that the degree of the circular polarization for optical transition
at each $k$-space point acquires a global minus sign, so that we
expect the opposite transition selection rule in the two $k$-space
valleys. Nevertheless, note that both the orbital magnetic moment
and the Berry curvature acquires a global minus sign as well when
$\Delta$ change sign. Thus, the optical selection rule always has
the same correlation with the valley contrasted magnetic moment and
the topological transport, as evident from Eq.~(\ref{curvature2}).
These physical properties in fact provide a better index for the two
inequivalent valleys than the $k$-space positions.

\section{Valley optoelectronics in graphene}
\label{section_application}

\begin{figure}
\includegraphics[width=7.5cm]{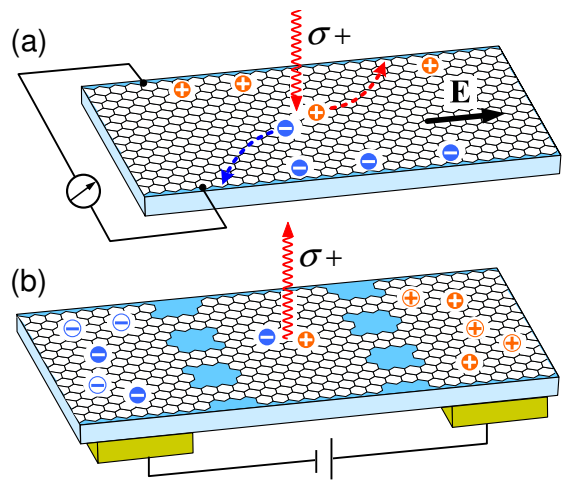}
\caption{(Color online) Schematic device geometry of optoelectronics
based on valley contrasting optical properties of graphene. The
electrons (holes) in valley $\rm{K}_2$ are denoted by white `-'
(`+') symbol in dark circles and their counterparts in valley
$\rm{K}_1$ are denoted by inverse color. (a) Photoinduced anomalous
Hall effect. (b) Valley light emitting diode. See text for the
explanation of operation mechanisms. } \label{device}
\end{figure}

Valley optical selection rule makes possible distinct electronic
response to light of different polarization, which maybe used as
potential principle of light polarimetry. We give an exemplary case
below exploiting the topological transport properties in graphene.
Inversion symmetry breaking in graphene leads to opposite Berry
curvatures distribution in the two valleys while the electrons and
holes at the same $k$-point have the identical Berry
curvatures.~\cite{XYN2007,particle-hole} In absence of magnetic
field, the net charge Hall current at equilibrium is zero as the
Hall flows at the two valleys exactly
cancel.~\cite{XYN2007,semenoff1984} Under the excitation by an
optical field with $\sigma-$ ($\sigma+$) polarizations, additional
electrons and holes are generated in valley $\rm{K}_1$ ($\rm{K}_2$)
and they move to opposite transverse edges of the sample if the
in-plane electric field is strong enough to dissociate the
electron-hole pairing (see Fig.~\ref{device}(a) for an
illustration). The sign of the developed transverse voltage thus
reflects the light polarization. For band-edge excitation, the
photo-induced Hall conductivity in the clean limit is $\sigma_H=\pm
4 \delta n \Omega_0 e^2/ \hbar $ for $\sigma \pm$ polarized light,
where $\Omega_0=2 \hbar^2 v_0^2 /\Delta^2$ is the Berry curvature at
the bottom of valley $\rm{K}_1$ and $\delta n$ is the density of the
photo-induced valley polarized electrons/holes. In the presence of
disorder, carriers may acquire an anomalous coordinate shift
proportional to the Berry curvature when they scatter off an
impurity potential.~\cite{Sidejump_smooth} This leads to a side-jump
contribution to the Hall conductivity which is also independent of
the scattering rate. The total Hall conductivity may have a sign
change as a function of the carrier
density.~\cite{XYN2007,Optical_dressing} Nevertheless, even in the
presence of disorders, the photo-induced Hall conductivity shall
ONLY depend on the photo-induced carrier density and the magnitude
of the on-site energy difference $|\Delta|$. The sign and size of
the Hall voltage induced by light of certain polarization are
independent of the density and details of the
disorders,~\cite{Sidejump_smooth} and are also independent of the
sign of $\Delta$. The latter is because the relative sign between
the Berry curvature and the degree of circular polarization is
universal (see Eq.~(\ref{curvature2}) and also the discussion in
section \ref{section_rule}).

Complementarily, control of luminescence polarization by electrical
means becomes possible in graphene. Mid-infrared light emitting
diode (LED) with electrically controlled emission polarization can
be realized in a similar fashion to the spin LED for
spintronics.~\cite{wolf2001} One possible configuration is shown in
Fig.~\ref{device}(b). The $n$-type and $p$-type regions are
connected to the central intrinsic region through nano-ribbons with
zigzag edges carved on the graphene sheet. These nano-ribbons act as
valley filters which preferentially allow right (left) moving
electrons and left (right) moving holes in valley $\rm{K}_2$
($\rm{K}_1$).~\cite{Beenakker_valleytronics1} For example, under the
applied gates bias shown in Fig.~\ref{device}(b), electrons
tunneling leftward from the $n$-type region and holes tunneling
rightward from the $p$-type region both have their majority
population in the $\rm{K}_2$ valley. In the central region,
electrons and holes bound to form excitons, and radiative
recombinations of these valley polarized excitons emit $\sigma+$
polarized mid-infrared photons. For SiC substrate with a dielectric
constant of $10$ and the resultant graphene energy gap
$\Delta=0.28$~eV,~\cite{grapheneGap_Lanzara} the radiative
recombination time for free excitons in graphene is estimated to be
$T_R \sim 10$~ps,~\cite{QWexcitonT1_1,QWexcitonT1_2} much shorter
than the observed intervalley scattering time $T_v \sim
100$~ps.~\cite{Gorbachev_WL_in_BG} Reversing the gates bias leads to
tunneling and recombination of electrons and holes dominantly from
valley $\rm{K}_1$ and the photon emitted from the central region
will have $\sigma-$ polarization. For small carrier density, both
the valley filtering effects and optical selection rules are nearly
perfect. Since two intervalley scatterings are needed to bring a
bright exciton from one valley to the other, we expect a
polarization loss of $(T_R/T_v)^2$ by intervalley scattering.

The electrically tunable energy gap in biased graphene bilayer is a
highly desirable property for optoelectronics as it enables the
interplay with light of a range of frequency. At zero doping, the
interband transition between the lower conduction band and upper
valance band may be implemented for a valley LED with an
electrically tunable emitting frequency. In metallic samples, the
transition between the two conduction bands is of interest, with a
peaked oscillator strength and perfect selection rules in the
vicinity of the Dirac points [Fig.~\ref{optical}(b,c)]. The Mexican
hat like dispersion in the lower conduction band further enables the
frequency selectivity. Thus, photo-induced anomalous Hall effect and
valley LED may also be realized in biased graphene bilayer with a
large sheet density by implementing this transition between the two
conduction bands.

\section{Conclusions}

A valley contrasting optical transition selection rule by inversion
symmetry breaking is demonstrated in the exemplary system of
graphene. In analogy to the spin degrees of freedom in
semiconductors, the valley index in graphene distinguishes the two
groups of electrons in their response to light with different
circular polarizations. Hence, besides the magneto optical
activities being
found,~\cite{GrapheneConductivity_Falko,magneto_optics,Jiang_magneto_optics}
graphene is of rich natural optical activities for technological
interests.~\cite{Wang_grapheneOptics} We show the possibility of the
valley analog of spin optoelectronics in graphene. Photo-induced
anomalous Hall effect~\cite{LightInducedHall,Optical_dressing} is
discussed as an example of converting light polarization information
into electronic signal. Complementarily, polarization of light
emission may be controlled electronically. We propose a graphene
based valley LED in the mid-infrared regime, with tunability in
emission frequency if realized in biased graphene bilayer. With the
usage of valley index as information carrier promised by the
inefficient intervalley
scattering,~\cite{Beenakker_valleytronics1,intervalley_theory1,intervalley_theory2,intervalley_theory3}
the valley selection rule can be exploited for a general class of
optically controlled graphene based logic with schemes borrowed from
optical control of spin information
processing.~\cite{wolf2001,spinQIP} In epitaxially grown graphene,
the as-prepared samples typically have a large sheet density $n\sim
10^{12} - 10^{13}$ cm$^{-2}$.~\cite{ohta2006,grapheneGap_Lanzara}
Adsorption of atom or molecular
acceptors~\cite{ohta2006,Wehling2007} can be used in combination
with gate voltage
control~\cite{novoselov2004,chiral_QHE1,chiral_QHE2} for
applications desired in the semiconducting regime.

This work is supported by the Welch Foundation, NSF under grant No.
DMR-0404252/0606485, DOE under grant No. DE-FG03-02ER45958, and NSFC
under grant No. 10740420252.

\begin{appendix}

\section{Dichroic sum rules} \label{appen_sum}

Overall circular dichroism has been suggested as a probe for the
orbital part of the magnetization in the
literature.~\cite{Smith_sumrules,Souza_Dichroism,THOLE_sumrules,Altarelli_sumrules}
When the orbital magnetism in solids is atomic like in nature, a sum
rule relates the integral of the circular dichroism to the value of
the orbital magnetization in ferromagnetic
systems.~\cite{THOLE_sumrules,Altarelli_sumrules} In insulators, in
the presence of intercellular current circulations, it is noticed
that the circular dichroism is only related to a portion of the
total orbital magnetization.~\cite{Souza_Dichroism} The universal
connection between the $k$-resolved optical oscillator strength and
the orbital magnetic moment shown in section \ref{sec_general}
indicates that, in both insulators and metals, the overall interband
circular dichroism shall reflect the magnetization contributed from
the band orbital magnetic moment. When the contribution comes from a
single band, we clearly see from Eq.~(\ref{obitalmoment}) and
(\ref{difference}) that
\begin{equation}
\frac{\mu_B}{2}(\langle f_- \rangle  - \langle  f_+ \rangle)  =
\hat{\bm z} \bm{\cdot} \int_{BZ} \frac{d\bm{k}}{(2\pi)^d} g( \bm k)
\bm{m} (\bm k) \label{f-sum},
\end{equation}
where $g(\bm k)$ is the Fermi distribution function. The right hand
side is simply the sum of the orbital magnetic moment of the filled
states, which constitute a gauge invariant contribution to the
orbital
magnetization.~\cite{orbitalmagnetization1,orbitalmagnetization2}
$\langle f_{\pm} \rangle$ stands for the sum of the interband
oscillator strength with $\sigma \pm$ polarized light
respectively~\cite{Yu_Cardona}
\begin{equation}
\langle f_{\pm} \rangle \equiv \sum_i \int_{BZ}
\frac{d\bm{k}}{(2\pi)^d} g( \bm k) \frac{\left|\mathcal{P}^{ci}_x
(\bm k) \pm \mathcal{P}^{ci}_y (\bm k)\right|^2}{m_e
\left(\varepsilon_c(\bm{k}) - \varepsilon_i(\bm{k}) \right)}.
\end{equation}
The total orbital magnetization also includes an additional gauge
invariant correction from the Berry phase
effect,~\cite{orbitalmagnetization1} which is not directly related
to the circular dichroism. The sum rule revealed here clearly shows
the physical significance of dividing the orbital magnetization into
these two gauge invariant portions.

The sum of the Berry curvature of the filled states constitutes the
intrinsic (clean limit) contribution to the anomalous Hall
conductivity in ferromagnetic systems which, in 2-dimention, is
given by~\cite{AHE_Niu}
\begin{equation}\sigma_H =
\frac{2e^2}{\hbar} \int_{BZ} \frac{d\bm{k}}{(2\pi)^2} g( \bm k)
\bm{\Omega}(\bm k) \bm{\cdot} \hat{\bm z}.
\end{equation}
The connection between the $k$-resolved optical oscillator strength
of interband transitions and the Berry curvature makes possible
optical measurement of this intrinsic contribution to anomalous Hall
conductivity. From Eq.~(\ref{curvature}), we find
\begin{equation}
\sigma_H = \frac{\epsilon_0}{\pi} \int d\omega (\epsilon^i_-(\omega)
- \epsilon^i_+(\omega) ) \label{f-sum2},
\end{equation}
where $\epsilon^i_{\pm}(\omega)$ is the imaginary part of the
dielectric function due to interband absorptions
\begin{widetext}
\begin{eqnarray}
\epsilon^i_{\pm}(\omega) = \frac{\pi e^2}{ \epsilon_0 m_e^2 \omega^2
} \sum_i \int_{BZ} \frac{d\bm{k}}{(2\pi)^2}  g( \bm k)
\left|\mathcal{P}^{ci}_{x} (\bm k) \pm i \mathcal{P}^{ci}_{y}(\bm
k)\right|^2 \delta(\varepsilon_c(\bm{k}) -
\varepsilon_i(\bm{k})-\hbar\omega).
\end{eqnarray}
\end{widetext}
Eq.~(\ref{f-sum2}) can be viewed as a manifestation of the
Kramers-Kronig relation on the interband part of the Hall
conductivity.~\cite{AHE_Niu}

\end{appendix}


\begin{thebibliography}{33}
\bibitem{nondegenerate}
This is the situation for the $s$-like conduction band in the
neighborhood of the $\Gamma$ point in III-V compounds, and the
active $\pi$ bands in graphene originated from $p_z$ atomic orbit of
carbon which has zero moment along the normal direction of the
plane. Here we do not consider the degeneracy from spin, which is a
dummy index in the limit of weak spin-orbit coupling. When the
parent atomic orbits do contribute to the orbital magnetic moment,
the Bloch bands are normally degenerate, and with large spin-orbit
coupling.
\bibitem{moment}
From Eq.~(\ref{ratio}), the selection rule at Dirac points implies
that $m(\bm{K}_{1,2})$ is a single quantum of
$\mu_B^{\ast}(\bm{K}_{1,2})$. In graphene single layer with a gap of
$0.3$~eV, the effective mass at Dirac points is $0.03 m_e$ and
therefore $\mu_B^{\ast} \sim 30 \mu_B$.
\end{thebibliography}
\end{document}